\documentclass[
 aip,
 jmp,
 amsmath,amssymb,
 reprint,%
]{revtex4-1}

\usepackage{graphicx}
\usepackage{dcolumn}
\usepackage{bm}

\begin{document}

\preprint{AIP/123-QED}

\title[Sample title]{Calibration System with Cryogenically-Cooled Loads \\
for CMB Polarization Detectors}

\author{M.~Hasegawa}
\author{O.~Tajima}
\affiliation{Institute of Particle and Nuclear Studies, High Energy Accelerator Research Organization~(KEK), Oho, Tsukuba, Ibaraki, 305-0801, Japan }
\author{Y.~Chinone}
\affiliation{Astronomical Institute, Graduate School of Science, Tohoku University, Aramaki, Aoba, Sendai, 980-8578, Japan }
\author{M.~Hazumi}
\author{K.~Ishidoshiro}
\author{M.~Nagai}
\affiliation{Institute of Particle and Nuclear Studies, High Energy Accelerator Research Organization~(KEK), Oho, Tsukuba, Ibaraki, 305-0801, Japan }

\date{\today}
             
\begin{abstract}
We present a novel system to calibrate millimeter-wave polarimeters for
CMB polarization measurements.
This technique is an extension of the conventional metal mirror rotation approach, 
however it employs cryogenically-cooled blackbody absorbers.
The primary advantage of this system is that it can generate a slightly polarized signal ($\sim100$~mK) in the laboratory; 
this is at a similar level to that measured by ground-based CMB polarization experiments observing a $\sim$ 10~K sky.
It is important to reproduce the observing condition in the laboratry for reliable characterization of polarimeters before deployment. 
In this paper, we present the design and principle of the system, and demonstrate its 
use with a coherent-type polarimeter used for an actual CMB polarization experiment.
This technique can also be applied to incoherent-type polarimeters 
and it is very promising for the next-generation CMB polarization experiments.
\end{abstract}


\maketitle

\section{Introduction}
\label{intro}

The faint pattern of cosmic microwave background~(CMB) polarization promises to provide 
new and interesting information on the universe.
The primary goal of CMB polarization studies in the next decade is to detect
the degree-scale $B$-modes~(curl components) induced by primordial gravitational waves.
The existence of the primordial gravitational waves is a generic prediction of inflation.
Therefore the detection of the $B$-modes is a ``smoking-gun'' signature of inflation~\cite{Zaldarriaga:1996xe}. 
Using an array with $\sim1,000$ of polarization detectors (hereafter polarimeters) 
is essential to discover the $B$-modes~\cite{Bock:2006yf}.
For the next-generation experiments with such a large polarimeter array,
establishing the performance of the detectors in the laboratory is essential to prepare for observations in the field.

Our system aims to calibrate three important properties; 
responsivity, orientation of detector polarization angle
and spurious polarization signal in the instrument.
The calibration of these parameters requires a well-characterized artificial polarization signal.
The combination of an unpolarized cold load (blackbody absorber) and a metal mirror at 
room temperature is one of the main approaches~\cite{Staggs:2006yf, kv:2008, Bischoff:2008wa} to obtain such a signal.
The temperature difference between the cold load and the mirror creates a
well-characterized polarization signal, $\sim100$~mK (Details are discussed in Sec.~\ref{chap:principle}).
In the laboratory, the blackbody absorber tends to be cooled with liquid nitrogen~($77$~K).
However, the large difference between this temperature and that of the sky at the observing site 
($\sim10$~K, for example the Atacama desert of northern Chile at an altitude of $\sim$~5000~m) 
makes it difficult to characterize the polarimeters in the laboratory.
This is particularly true for detectors with a narrow range of linear response to input load temperature.

To address these issues, we propose an advanced
approach for polarimeter calibration. 
Our method is an extension of the metal mirror rotation
approach~\cite{Staggs:2006yf, kv:2008, Bischoff:2008wa}, 
however it employs cryogenically-cooled blackbody absorbers 
instead of the absorbers cooled with liquid nitrogen.  
This system provides in the laboratory 
a load condition similar to the actual observation.

\section{Principle of Polarization Calibrator}
\label{System_Overview}

\subsection{Principle of polarization signal generation}
\label{chap:principle}

\begin{figure*}[hbt]
  \begin{tabular}{cc}

    \begin{minipage}{0.45\hsize}
      \vspace{0.0cm}
       \begin{center}
          \resizebox{7.5cm}{!}{
           \includegraphics[trim=0cm 0.0cm 0.0cm 0cm, clip]{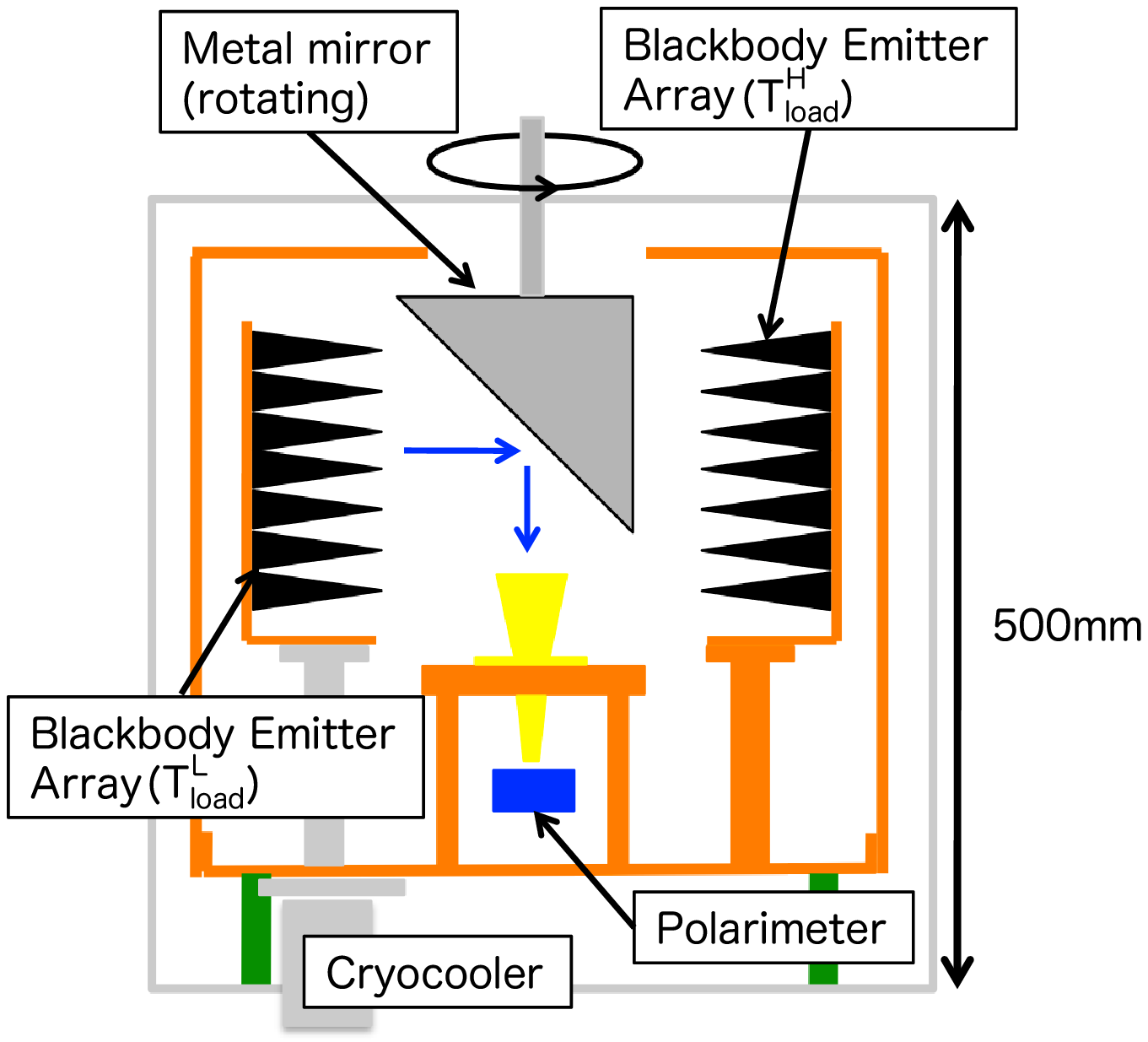}}       
      \end{center}
   \end{minipage}   

    \begin{minipage}{0.15\hsize}
    \end{minipage}     

     \begin{minipage}{0.45\hsize}
       \begin{center}
         \resizebox{6.5cm}{!}{
         \includegraphics[trim=0.0cm 0.0cm 0.0cm 0cm, clip]{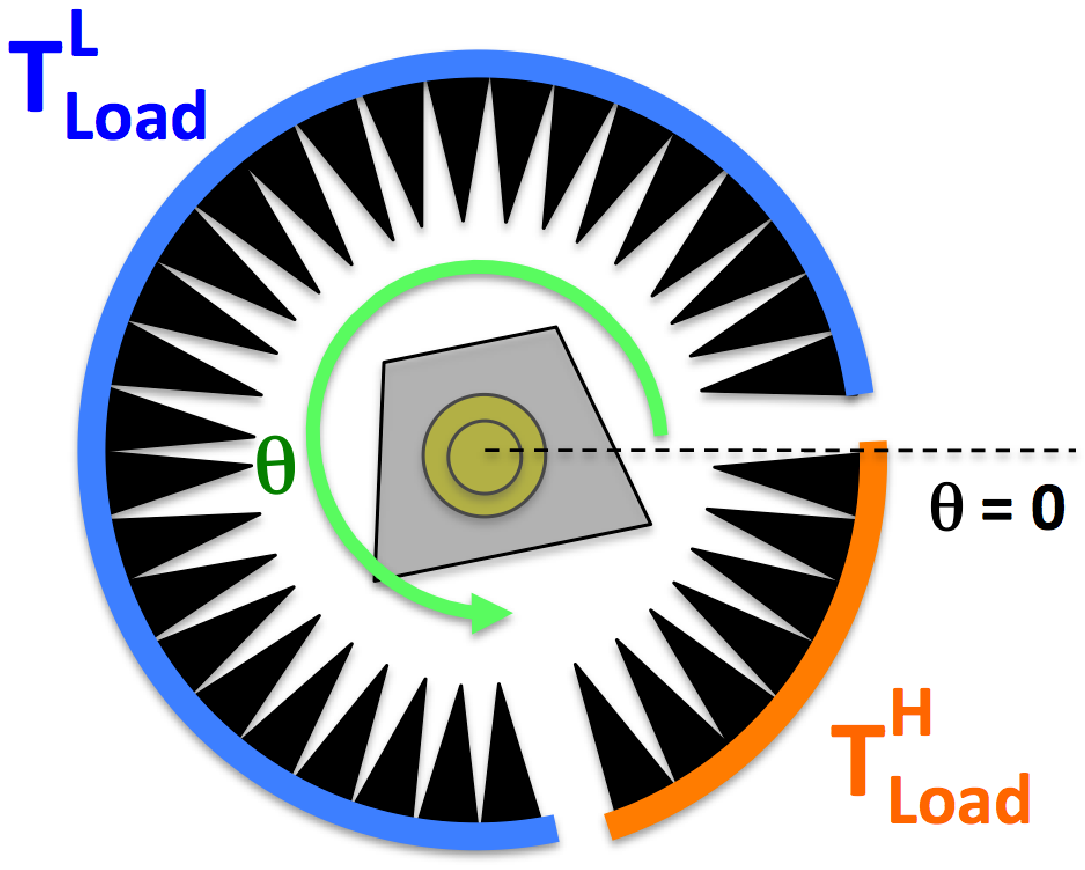}}
      \end{center}
    \end{minipage}   
 \end{tabular}
   \caption{ Schematic view of the developed calibration system. Unpolarized radiation is emitted from the blackbody absorbers
             cooled with a cryocooler. The radiation reflects off a metal mirror surface placed at the center, 
             and enters a tested polarimeter. The incident radiation is polarized because of the finite emissivity of the mirror.
             The blackbody absorbers are cylindrically placed as shown to the right.
             The cryocooler in the system has two 
             cooling stages, and a part of absorbers are cooled to be different~(higher) temperature using another cooling stage. 
             The mirror can rotate in the vacuum, and the loads at two different temperatures are alternately presented to the polarimeter.}
   \label{fig:schem2}
 \end{figure*}  

The principle for generating the polarization signal is the same as the conventional method~\cite{Staggs:2006yf, kv:2008, Bischoff:2008wa}. 
Figure~\ref{fig:schem2} shows a schematic view of the developed calibrator system.
The blackbody emitters are cooled with a cryocooler as the cold load with temperature of $T_{\rm load}$. 
 Unpolarized blackbody radiation is emitted from the cold load; it reflects off a metal mirror surface, which induces
a linearly polarized component because of the finite emissivity of the mirror. 
The magnitude of the polarization signal ($P$) can be calculated 
using the following formula,
\begin{eqnarray}
P  &=&  2\sqrt{16\pi\nu\rho\varepsilon_{0}}\,\tan\left( \beta \right) \left( T_{\rm mirror} - T_{\rm load} \right), \label{eq:pol}
\end{eqnarray}
where $\nu$ is the frequency of observed radiation, 
$T_{\rm mirror}$ and $\rho$ are 
the temperature and resistivity of the mirror, 
$\beta$ is the reflection angle of the radiation, 
and $\epsilon_0$ is the vacuum permittivity.
For example, we obtain $\sim$110~mK polarization signal when we use an aluminum mirror~($\rho \sim 2.2~\mu\Omega\cdot \rm{cm}$) 
with $T_{\rm mirror}=250$~K, $T_{\rm load}=10$~K and $\beta = 45^{\circ}$.
We can control the signal magnitude by changing $\rho$ as well as changing the temperature difference, $T_{\rm mirror} - T_{\rm load}$.
%

\subsection{Principle of Polarimeter Calibration}
\label{principle_of_calibration}

The linear polarization is characterized 
by two Stokes parameters $Q$ and $U$~\cite{Zaldarriaga:1996xe}. 
Since the value of Stokes parameters 
depends on the bases of the selected coordinate system,  
the ability to rotate the axis of the polarization signal is essential to calibrate the polarimeter responses
unless we calibrate the orientation of the polarization angle for the Stokes parameters.

Moreover, the baseline of the polarimeter response fluctuates due to a $1/f$ noise.
To measure the response for the polarization correctly, we have to modulate the polarization signal by rotating the mirror;
the parallactic angle of the signal axis with respect to the polarimeter varies with the rotation angle of the mirror ($\theta$). 
For example, assuming the axis for the $Q$ response of the polarimeter aligns with the orientation of $\theta$ = 0, 
the $Q$ response varies with $P\cos(2\theta)$,
and the $U$ response varies with $P\sin(2\theta)$~$\left(= P\cos(2(\theta - \frac{\pi}{4})) \right)$.
We can characterize the responsivity; signal height with respect to the input polarization signal, 
and the polarization angle; orientation angle of the polarization axis 
from the amplitude and the phase of the sinusoidal response of the polarimeter, respectively.

We are also interested in the spurious polarization of the polarimeter with respect to the total power 
which is usually dominated by unpolarized temperature.
This is called $I$ to $Q$ (or $U$) leakage where the total power is described with Stokes $I$ parameter.
The actual polarization response for $Q$ is $P\cos(2\theta) + \varepsilon I$, where $\varepsilon$ is the leakage parameter 
Since $I$ is dominated by $T_{\rm load}$ in most cases, 
we can measure the leakage parameter for the polarimeter by changing the $T_{\rm load}$~\cite{A_comment}.   
With the two different temperature loads of 10~K and 30~K, 
we observe a 1~K baseline difference in the polarization response for $\varepsilon =0.05$. 

The polarimeter can also measure the unpolarized temperature 
response for $I$ as well as for $Q$ and $U$.
We calibrate the response for $I$ by using two different $T_{\rm load}$~s (the Y-factor measurement).
This allows us to measure receiver temperature~($T_{\rm rec}$) which is the effective temperature 
offset from the intrinsic noise.

\subsection{Expected Polarimeter Response }
\label{expected_response}

\begin{figure}[tbh]

      \vspace{0.0cm}
       \begin{center}
          \resizebox{9.5cm}{!}{
          \includegraphics[trim=0cm 0.0cm 0.0cm 0cm, clip]{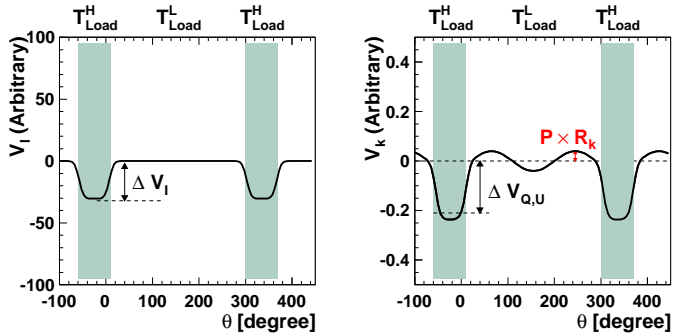}}
      \end{center}
 
   \caption{The expected response 
   of total power and polarization signal plotted as a function of the rotation angle~($\theta$). 
   The two different temperature pyramid arrays are alternately viewed by the rotating metal mirror.     }
    \label{fig:SigExp}
 \end{figure}  

In general, the polarization angle does not align with the design value perfectly.
The responses of the polarimeter are 
defined as described in the following formula in the unit of mV,
\begin{eqnarray}
V_Q &\equiv&  R_Q 
                                ~\left[~
                                P\cos\left( 2(\theta - \phi_Q) \right) 
                                \right.
                                \nonumber \\
                                && \hspace{4.5em} 
                                \left.
                                        + \varepsilon_Q (T_{\rm load}+P)
                                        ~\right]
                                        , 
                                        \label{eq:V_Q}
\\
V_U &\equiv&  R_U 
                                ~\left[~
                                P\sin\left( 2(\theta - \phi_U) \right) 
                                \right.
                                \nonumber \\
                                && \hspace{4.5em} 
                                \left.                          
                                + \varepsilon_U (T_{\rm load} + P)
                                ~\right]                                        
                                , 
                                \label{eq:V_U}
\\
V_I &\equiv&  R_{I~} ~\left[~T_{\rm load} + P +  T_{\rm rec} ~\right] ,
                                        \label{eq:V_I}
\end{eqnarray}
where $R_{k}~(k= Q, U)$ and $R_{I}$ are the responsivities (mV/K),
$\phi_{k}$ is the polarization angle 
and $\varepsilon_{k}$ is the leakage parameter
for each Stokes parameter response.

Figure~\ref{fig:SigExp} shows 
the polarimeter responses as a function of the rotation angle of the mirror.
Here, we assume 
a configuration with two different temperature loads $T_{\rm load}^{\rm L} = 10$~K and $T_{\rm load}^{\rm H} = 30$~K
and the leakage parameter $\varepsilon = 0.05$.
The $T_{\rm load}^{\rm L}$ load covers the $10^{\circ} \sim 300^{\circ}$ region and other regions are covered by the $T_{\rm load}^{\rm H}$ load.
The boundary regions between the two loads are masked in the analysis 
because it is complicated to estimate the response due to the beam profile.

We can extract all the calibration parameters from the simultaneous fit for each polarimeter response with respect to the mirror rotation angle ($\theta$) and the load temperatures ($T_{\rm load}$).
In the case   
$T_{\rm load}$ is constant, the polarimeter responses $V_{k}$ are simple sinusoidal shape with constant offsets.
Furthermore, in the case $\varepsilon_{k} \Delta T_{\rm load} \gg P$ (where $\Delta T_{\rm load} \equiv T_{\rm load}^{\rm H} - T_{\rm load}^{\rm L}$),
the leakage parameters $\varepsilon_{k}$ can be extracted from the baseline difference
$\Delta V_{k}$ between $T_{\rm load}^{\rm H}$ and $T_{\rm load}^{\rm L}$ , i.e. 
$\varepsilon_{k} = \Delta V_{k} / (R_{k} \Delta T_{\rm load})$.

\section{System Components}

\subsection{Cryostat and Cryocooler}

A cylindrical cryostat 
with a dimension of 540~mm diameter and 500~mm height 
houses all of the system components as shown in Fig.~\ref{fig:schem2}.
The system is cooled
with the 2-stage Gifford-McMahon cryocooler~\cite{sumitomo}
Its cooling capacity in the cryostat 
is consistent with the specification~(Appendix \ref{loadmap}).
Thermal load for the 1st (2nd) stage in our system is about 18~W (7~W), and
we usually operate at 30~K (10~K) temperature on the top of the 1st (2nd) stage.

\subsection{Cold Load -- Blackbody Emitter Array}

\begin{figure}[tbph]
        \begin{center}
                \resizebox{5.5cm}{!}{
                \includegraphics[trim=0cm 0cm 0cm 0cm,clip]{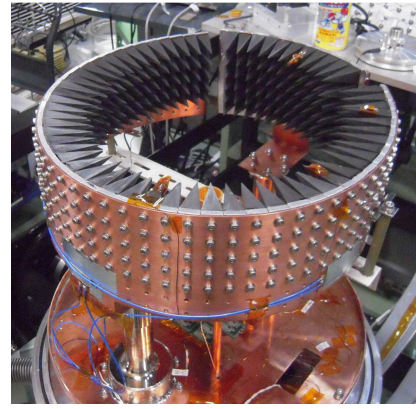}}              
                \caption{CR-112 blackbody emitter array as the cold load.}
                \label{fig:pyramid_array}
        \end{center}
\end{figure}
 
 \begin{figure}[tbph]
        \begin{center}
                \resizebox{8.5cm}{!}{
                \includegraphics[trim=0.0cm 0cm 0cm 0cm,clip]{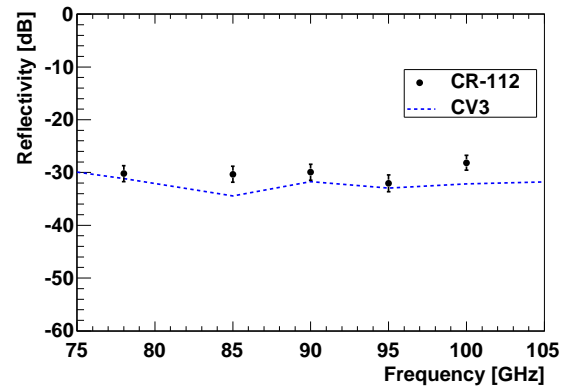}}             
     \caption{Measured power reflection coefficient of CR-112 and CV-3. }
    \label{fig:reflectivity}
\end{center}    
\end{figure}

To create the low temperature blackbody radiation, 
low reflectivity and high thermal conductivity are required 
for the blackbody emitters.
Furthermore, the emitters need to have viscosity (or similar thermal expansion 
coefficients with the base plate material) 
to withstand the thermal cycling.
To fulfill the requirements, we employ Eccosorb CR-112~\cite{eac}
an iron-loaded epoxy 
as the material for the blackbody emitter.
To minimize the surface reflection, 
the CR-112 is casted with pyramid-shape grooves on the front surface. 
We follow the manufacturing processes of the ARCADE experiment~\cite{Kogut:2004hq}.
The pyramid has a square base of 400~mm$^2$ and height of 58~mm. 
To achieve higher thermal conductivity and uniform surface temperature, 
each pyramid has an aluminum core 
and a copper wire epoxied onto the end of the aluminum core, running almost to the tip.
We use two arrays of the pyramids as the cold load (Fig.~\ref{fig:pyramid_array}).
We use 420 pyramids in total to surround the inside of the cryostat wall in the system.

We confirmed the low reflectivity of the pyramid array 
by measurements with 
commercial millimeter-wave equipment~\cite{equipment_reflection}.
The measured reflectivity for the pyramid array is shown in Fig.~\ref{fig:reflectivity}. 
It is about $-30$~dB around the 90~GHz frequency region and is comparable to commercial 
absorber material such as CV-3$^{8}$. 

\subsection{Metal Mirror}
\label{mirror}

\begin{table*}
 \begin{center}
  \caption{  Resistivity ($\rho$), load temperatures and mirror temperature for each mirror material.
  $P$ is the calculated polarization amplitude which corresponds to each $T_{\rm load}^{\rm L}$.
  In case of the stainless steel mirror, we use an aluminum shaft, which has better heat conductivity than that of 
  a stainless steel shaft.  
  Therefore $T_{\rm mirror}$ is higher than others.
  } 
    \label{tab:mirror}
    \begin{tabular}{l c c c c c}
     \hline
     \hline
       Mirror Material  
       & \hspace{1mm} $\rho$~[$\mu\Omega\cdot\rm{cm}$ ] \hspace{1mm} 
       & \hspace{1mm} $T_{\rm load}^{\rm L}$~[K] \hspace{1mm} 
       & \hspace{1mm} $T_{\rm load}^{\rm H}$~[K] \hspace{1mm} 
       & \hspace{1mm} $T_{\rm mirror}$~[K] \hspace{1mm} 
       & \hspace{1mm} $P$~[mK] \cr
     \hline
        Aluminum~(6061)           &   ~2.2    & 12.0   & 29.7  &  250.0 & 114.8  \cr
        Steel   ~(304)            &   17.1    & 13.5   & 30.0  &  250.0 & 370.0  \cr
        Stainless Steel~(1095)    &   54.9    & 14.5   & 32.1  &  257.2 & 591.9  \cr
     \hline
     \hline
   \end{tabular}   
  \end{center}
\end{table*}      

We adopt $\beta = 45^{\circ}$ 
for the metal mirror in our system.
The mirror is hung with a stainless steel shaft that
is rotated with a DC motor from the outside of the cryostat via a feedthrough~\cite{CI}
The maximum rotation speed is $\sim$12~rpm,  
which corresponds to much higher frequency of the $1/f$ noise of the polarimeters for the CMB experiment (several mHz $\sim 100$~mHz).

We use three metal mirrors, aluminum, steel and stainless steel, to vary the polarization signal intensity. 
It is useful to evaluate the possible inherent polarization~\cite{kv:2008, bruce:2008}.
We obtained zero-consistent inherent polarization in our system (Sec.~\ref{chap:poloffset}).

Currently, precision of the resistivity ($\rho$) is the main uncertainty for the calculation of 
the polarization signal ($P$).
The variation of the catalog specification is 20\%, 
corresponding to a 10\% uncertainty for the polarization signal calculation.
Such uncertainty can be reduced with a resistivity measurement in a separate setup.

\subsection{Load and Metal Mirror Temperature}
\label{sec:temperature_achievements}

\begin{figure}[tbph]
        \begin{center}
                \resizebox{9cm}{!}{
                \includegraphics[trim=0cm 0cm 0cm 0cm,clip]{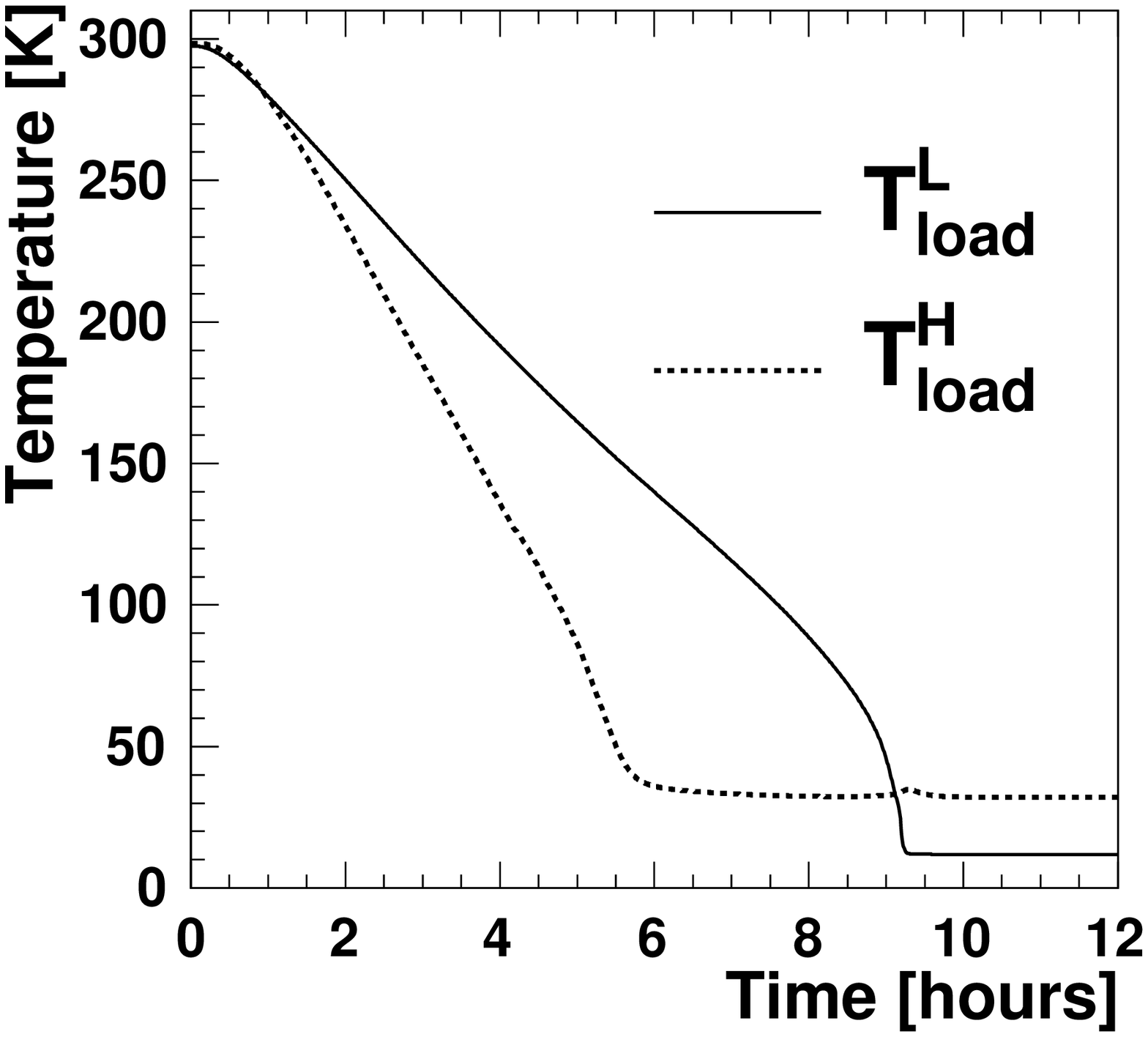}
                \includegraphics[trim=0cm 0cm 0cm 0cm,clip]{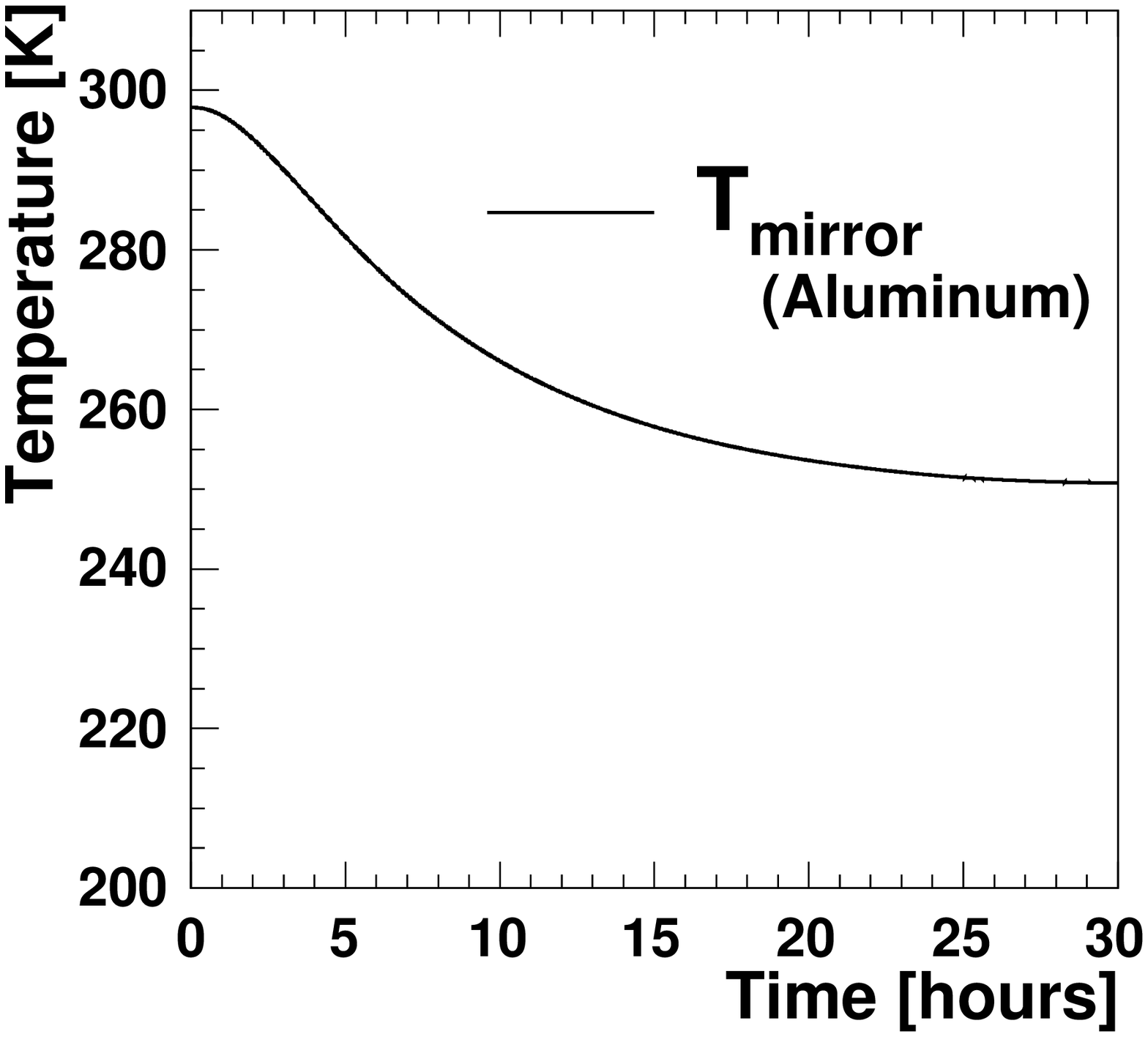}}
                \caption{Loads and mirror temperatures during the cooling process. }
                \label{fig:cooldown}
        \end{center}
\end{figure}

Figure~\ref{fig:cooldown} shows the temperature of the loads and the aluminum mirror.
It takes $\sim$10~($\sim$ 30) hours to stabilize the temperature of the cold load~(mirror)
Because of the difference of the emissivities for each metal, 
equilibrium temperatures depend on the mirror material as summarized in Table~\ref{tab:mirror}.
The measured temperatures are consistent with the thermal calculation in Appendix \ref{chap:heat}.

Uniformity of the load surface temperature is confirmed to be better than 1~K
with DT-670 commercial silicon diode thermometers~\cite{LakeShore}
The temperature non-uniformity causes only 
a 0.4\% effect for the polarization signal 
while the effect for the unpolarized response is about 10\%.
The temperature uniformity can 
also be constrained from the flatness of the $V_I$ response 
as a function of the rotation angle effectively, which is also less than 1~K (Sec.~\ref{sec:tp_response}).

\section{Evaluation of System Performance}

\subsection{Polarimeter for System Evaluation}
\label{sec:QUIET_polarimeter}

We evaluate the system performance
using the QUIET polarimeter~\cite{bib:QUIET-Module}.
QUIET has an array of pseudo-correlation polarimeters, each of them employs
hybrid couplers to correlate the two orthogonal polarizations selected by an orthomode transducer 
in order to directly and simultaneously measure the Stokes $Q$, $U$ and $I$ parameters of the incoming signal. 
A corrugated horn is attached on top of each polarimeter, and its beam width is measured to be $\sim 10^{\circ}$~(FWHM)~\cite{bib:Gundersen}.
The beam coverage of  this system is $\sim 20^{\circ}$, and the spill-over power is expected to be less than 1\%. 
The responsivity of this polarimeter is
$\sim 0.15\, \rm{mV/K}$ including the pre-amplifier gain factor of $\sim 100$~\cite{B_comment}
which was measured with 
a different setup using the rotation plate technique under 77~K.
In case of QUIET polarimeters, 
reflections around the septum polarizer is the main reason 
for the $I$ to $Q$ ($U$) leakage.
A fraction of the power input on one port of the correlation module was reflected to the septum polarizer and a fraction of the reflected power re-entered the other port.
Typically, the leakage of a QUIET polarimeter is $\varepsilon \sim 0.005$.
To evaluate the system performance for  the leakage parameter ($\varepsilon$) easily,
we insert an extra waveguide between the corrugated horn and the septum polarizer. 
This increases the reflection, and the leakage size is increased to 
$\varepsilon \sim 0.05$.

\begin{figure*}[tb]
  \begin{tabular}{cc}
     \begin{minipage}{0.45\hsize}
       \begin{center}
          \resizebox{8cm}{!}{
          \includegraphics[trim=0.0cm 0.0cm 0.0cm 0cm, clip]{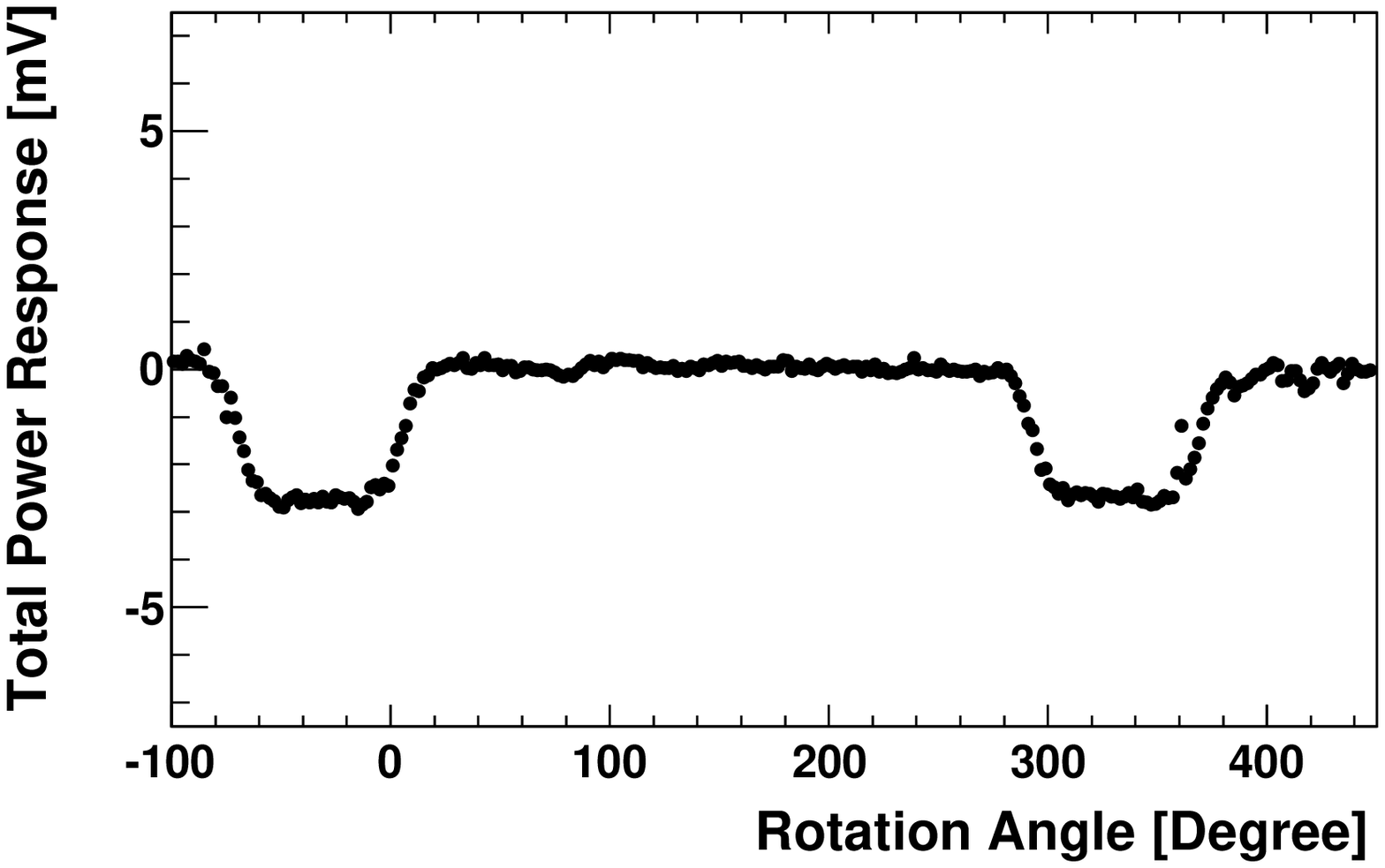}}
      \end{center}
   \end{minipage}   
   
    \begin{minipage}{0.50\hsize}
      \vspace{0.0cm}
       \begin{center}
          \resizebox{8cm}{!}{
          \includegraphics[trim=0cm 0.0cm 0.0cm 0cm, clip]{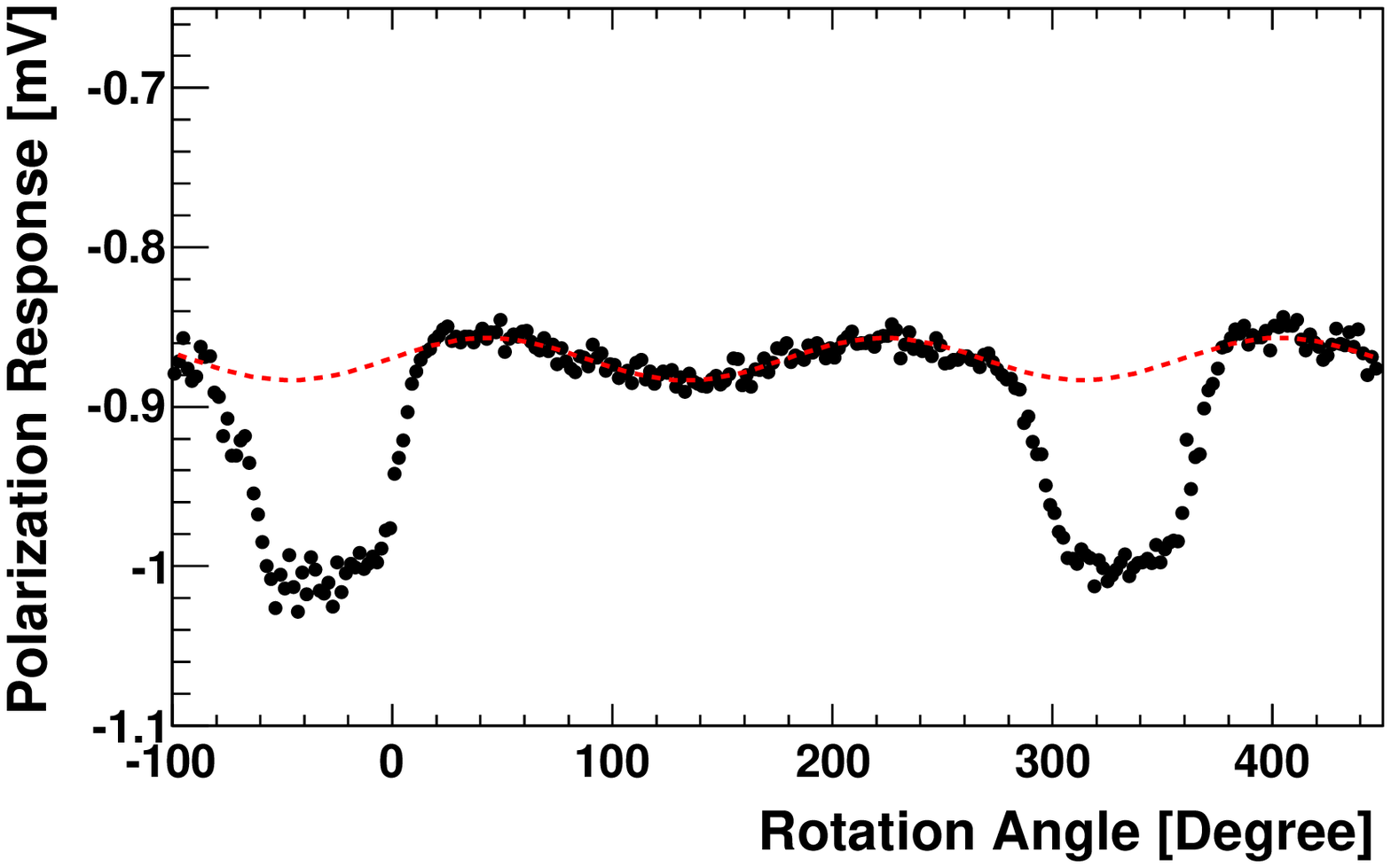}}
      \end{center}
   \end{minipage}
 \end{tabular}
   \caption{Response of the QUIET polarimeter to the rotating aluminum block. The left panel shows the timestream 
   for a total power measurement and the right panel shows the timestream 
   for a polarization measurement. A red dashed line in the right panel is a fitted sinusoidal curve.}
    \label{fig:TP}
 \end{figure*}

\subsection{Total power response}
\label{sec:tp_response}

The response for the total power as a function of the rotation angle of the mirror
is shown in the left panel of Fig.~\ref{fig:TP}.
The total power response shows two levels;
the upper~(lower) level corresponds to 10~K~(30~K) load temperature.
The total power responsivity, $R_{I}$, is calculated from the timestream 
with Eq.~\ref{eq:V_I} to be 0.15~(mV/K),
which is consistent with the previous measurement discussed in Sec.~\ref{sec:QUIET_polarimeter}.

The uncertainty for the total power response is dominated by the uniformity of the load temperature,
which is evaluated from the timestream of 
the total power response. 
The RMS of the total power response for the $T^{\rm L}_{\rm load}$ load is 0.9~K,
which is consistent with 
the RMS on the temperature monitored using thermometers.  
This variation includes the polarimeter noise which dominates the response fluctuation in this test.
Thus, this number still provides an upper limit on the non-uniformity of the load temperature.
So far, the precision for the total power responsivity measurement is 7\% 
(1~K uncertainty both for $T_{\rm load}^{\rm L}$ and $T_{\rm load}^{\rm H}$),
which is already comparable with the precision obtained using astronomical sources.

\subsection{Polarization response}

The response for the polarization as a function of 
rotation angle of the aluminum mirror is shown in the right panel of Fig.~\ref{fig:TP}.
The response shows a clear sinusoidal curve as a function of the mirror rotation angle.
This is the first demonstration~``in the laboratory''
for the modulation of the polarization signal (110~mK)
with a load temperature which is as low as those at the observation site.

From the amplitude of the curve, we obtain 
$R_U = 0.12 \pm 0.01 (stat) \pm 0.02 (syst)$~mV/K
which is consistent with the $R_I$.
Here the $stat$ term is the statistical uncertainty due to the polarimeter noise 
and the $syst$ term is the systematic uncertainty derived from the catalog uncertainty of the resistivity (10\%, Sec.~\ref{mirror}) and the load temperature (0.4\%).
This is the uncertainty for the absolute scale, not for the relative uncertainty among detectors.
The relative uncertainty is much more important for CMB polarimeters.

For the polarization angle, $\phi_{U}$, we extract the angle with a 0.8$^\circ$ error from the fit, 
where the error is dominated by the polarimeter noise. 
The system has not been limited by the precision of the relative angle measurement.
However, we have not 
implemented the encoder for the mirror rotation angle yet, which limits the absolute angle measurement.
This encoder is needed to achieve the necessary precision of 0.5$^{\circ}$
for future $B$-modes measurements~\cite{Hu:2002vu}.

For the $I$ to $Q$ (or $U$) leakage, we can determine the leakage parameter with 
a precision of 0.005 from Fig.~\ref{fig:TP}.
Again this measurement is limited by the polarimeter noise.
There is no systematic limitation for the leakage measurement 
unless the relative uncertainty between $R_I$ and $R_{Q,U}$ becomes significantly large.

Just a few tens of seconds of measurements using our system can provide 
comparable precision with astronomical source calibration at the observation site, for example 
a 20-minute observation for the Crab nebula~(Tau~A)~\cite{Bischoff:2010ud}. 
Our system therefore provides better precision than astronomical sources in a given amount of time.

\subsection{Evaluation of Inherent polarization signal in the system}
\label{chap:poloffset}

\begin{figure}[tbph]
        \begin{center}
                \resizebox{8cm}{!}{
                \includegraphics[width=9.5cm, height=6.0cm, trim=0cm 0cm 0cm 0cm,clip]{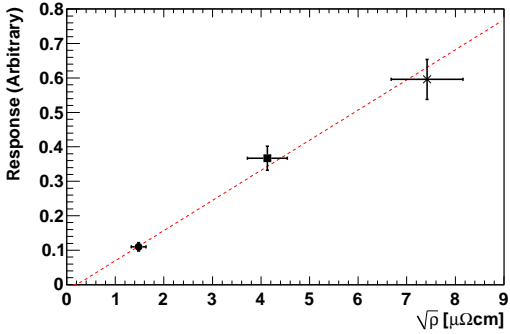}}                
                \caption{The response of the polarized signal is plotted with respect to the square root of 
                the resistivity of the mirror material. Here three different materials~(Aluminum~($\bullet$), 
                Steel~(${\blacksquare}$) and Stainless steel~($\ast$)) are used.
                     }
                \label{fig:response}
        \end{center}
\end{figure}

Figure~\ref{fig:response} shows the amplitude of 
the polarized signal for three different mirror materials.
The results show that the amplitudes are proportional 
to the square root of the resistivity as expected, and
the system has no inherent polarized signal within the precision of the measurement.
This is the evidence that there is no bias for the polarization responsivity measurement
by using the single mirror setup.

\section{Discussion and Summary}
\label{conclusion}

We have developed a polarization calibrator for CMB polarization receivers. 
The technique is a simple extension of the conventional rotation metal plate approach, 
however, it employs $\sim$10~K cold loads which correspond to the sky temperature at the CMB observation site on the ground (the Atacama desert of northern Chile at an altitude of $\sim$ 5000~m).
The calibrator can provide a well-characterized polarization signal ($\sim 100$~mK) 
which can be modulated by rotation of a mirror.
Using a QUIET polarimeter, we demonstrated 
simultaneous measurements of the responsivity, orientation of polarization angle, and spurious polarization signal in the instrument with sufficient precision.
These three parameters are keys to the success of 
$B$-mode observations as well as sensitivity~(noise equivalent temperature: $\mu {\rm K}\sqrt{{\rm s}}$) of the polarimeter.
Our system also provides a way to measure noise properties
under a comparable load temperature condition to the observing site.
The estimation based on our system is thus more ``reliable'' than 
that obtained using a 77~K load cooled with liquid nitrogen.
Our system can easily be extended for large detector arrays by scaling up the size of the system.

The precision of the laboratory-based calibration is comparable with the 
astronomical source calibrations at the observation site. 
This system can establish the polarimeter performance before deployment, 
which is important for experiments with large detector arrays.
Moreover, the precision of calibrations using astronomical sources is limited by the beam profile uncertainty.
Our system is free from this effect because of the wide coverage-angle of the loads. 
We have already eliminated it to less than 1\% 
for the QUIET polarimeter.

We have confirmed that our system works as designed. 
It will be very useful for the next-generation CMB polarization experiments.
In this paper, we demonstrated the performance of the system 
using a polarimeter based on coherent amplifiers.
By using a different configuration, e.g. a setup in which the reflected emission on the mirror goes outside the cryostat,
it is possible to use the system as an external polarized source for a polarimeter with different technology such as a bolometer.
With a larger-capacity cryocooler, the system can also be extended to 3~K load temperatures.
Such a system will be a key tool in the development of polarimeters for the future satellite experiments.

\section{Acknowledgement}
\label{Acknowledgement}

We acknowledge Todd Gaier, Kieran A. Cleary and their colleagues 
at Jet Propulsion Laboratory and California Institute of Technology for providing 
low-noise polarization sensitive modules and valuable advice on the calibration system. 
We would like to thank Michele Limon for valuable information about the cold load.
We also thank Hogan Nguyen, Fritz DeJongh, Akito Kusaka and Bruce Winstein for their encouragement. 
We wish to thank Takayuki Tomaru and their colleagues at Cryogenics Science Center 
for their cooperation and advice on the low temperature techniques. 
We gratefully acknowledge the cooperation of Bee Beans Technologies Co.,Ltd.
This work was supported by MEXT and JSPS with Grant-in-Aid for Scientific Research 
on Young Scientists B 21740205, Scientific Research A 20244041, and Innovative Areas 21111002.

\appendix

\section{Evaluation of Cooling Capacity of Cryocooler}
\label{loadmap}

\begin{figure}[tbph]
        \begin{center}
                \resizebox{8.5cm}{!}{
                \includegraphics[ trim=0cm 0cm 0cm 0cm,clip]{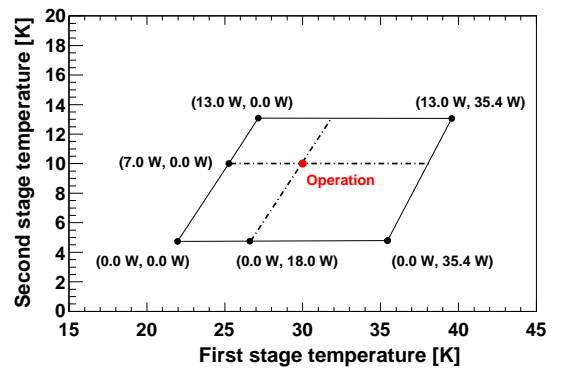}}           
                \caption{ Measured Load map of the cryocooler used in the system, SHI-RDK 408S, 
                          which is consistent with the sepcification.}
                \label{fig:loadmap}
        \end{center}
\end{figure}

Figure~\ref{fig:loadmap} shows the measured load map 
for the 10\,K Gifford-McMahon cryocooler~(SHI-RDK 408S)
used in the calibrator system.
The first~(second) stage of the refrigerator reaches temperature of 22\,K~(4.6\,K)
under no external thermal load.
The calibration system is usually operated at 
30\,K~(10\,K) on the first~(second) stage.
The thermal load on each stage during the operation 
is estimated to be 18\,W~(7\,W) from the load map measurement. 


\section{Heat transfer on the metal mirror and mirror temperature in the system}
\label{chap:heat}

\begin{figure}[tbph]
        \begin{center}
                \resizebox{6cm}{!}{
                \includegraphics[ trim=0cm 0cm 0cm 0cm,clip]{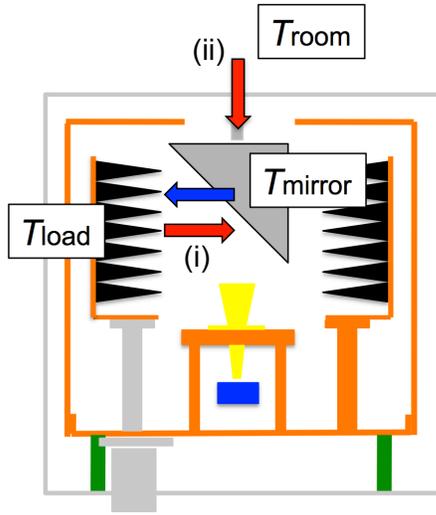}}              
                \caption{ Schematic diagram of a cryostat for heat transfer calculation on the mirror.}
                \label{fig:heatroad}
        \end{center}
\end{figure}

Figure~\ref{fig:heatroad} shows a schematic diagram of 
the heat transfer on the metal mirror. The mirror has a surface area of $0.04~\rm{m}^2$
and is supported by a stainless steel shaft~(the surface area divided by the length is 0.09~m).
The mirror temperature instationary state is estimated as the
temperature 
where the following heat transfer balances.
\begin{enumerate}
\item $\dot{Q}_{\rm{rad}}$ : Radiant heat inflow from the cold loads and outflow. 
\item $\dot{Q}_{\rm{con}}$ : Heat conduction from a DC motor through the shaft.  
\end{enumerate}

Each heat transfer is calculated as follows;
\begin{eqnarray}
\dot{Q}_{\rm{rad}} &=& \sigma~A_{\rm mirror} \left(  -T^{4}_{\rm mirror} + T^{4}_{\rm load} \right) \varepsilon_{\rm mirror}   \\
\dot{Q}_{\rm{con}} &=& \frac{A}{L}\int^{300{\rm K}}_{T_{\rm mirror}} k(T)dT
\end{eqnarray}
where $\sigma$ is Stefan's constant~($5.67\times10^{-8}\rm{W m}^{-2}\rm{K}^{-4}$),  $A_{\rm mirror}$ is the surface
area of the mirror, $T_{\rm mirror}$ and $T_{\rm load}$ are the physical temperature of the mirror and the cold load,
$\varepsilon_{\rm mirror}$ is the emissivity of the mirror,
$A/L$ is the ratio of the surface area to the length of the stainless steel shaft, and
$k(T)$ is the temperature-dependent thermal conductivity of the shaft, respectively.   
In case we use an aluminum mirror~($\varepsilon_{\rm mirror}$ $0.03 \sim 0.04$), 
the heat transfers are balanced at $T_{\rm mirror}$ of 240~K $\sim$ 260~K.
This is consistent with the measured temperature as stated in Sec.~\ref{sec:temperature_achievements}.


\begin{thebibliography}{18}
\bibitem [{A_c()}]{A_comment}%
  \BibitemOpen
  \href@noop {} {}\bibinfo {note} {In higher order calculation, the $I$ is
  define with sum of $T_{\rm load}$ and $P$. In case we use two different
  temperature loads, the difference of the load temperature and the
  polarization signal are defined with symbols $\Delta T_{\rm load}$ and
  $\Delta P$ respectively. Therefore, the variation of $I$ is described as
  $\Delta I \equiv \Delta T_{\rm load} + \Delta P$, and $\Delta I \sim \Delta
  T_{\rm load}$ in most of cases because $\Delta P \sim \Delta T_{\rm
  load}/100$}\BibitemShut {NoStop}%
\bibitem [{equ()}]{equipment_reflection}%
  \BibitemOpen
  \href@noop {} {}\bibinfo {note} {We obtained the signal in W-band region with
  Agilent E8247C (PSG) CW signal generator and frequency multiplier chain
  (AMC-10-RFH000). The W-band signal was injected to the blackbody emitter
  array with an incident angle of 7$^\circ$. The reflected signal power was
  measured by a general purpose detector (DXP-10 RPFW0).}\BibitemShut {Stop}%
\bibitem [{B_c()}]{B_comment}%
  \BibitemOpen
  \href@noop {} {}\bibinfo {note} {The typical responsivity of the QUIET
  polarimeters is 2 $\sim$ 3mV/K. The module used for the test in this paper is
  a low gain module which was not used for the CMB observation.}\BibitemShut
  {Stop}%
\end{thebibliography}

\providecommand{\noopsort}[1]{}\providecommand{\singleletter}[1]{#1}%

\end{document}